\documentclass[twocolumn, secnumarabic, amssymb, superscriptaddress, aps, prl, floatfix]{revtex4-1}
\usepackage{amsmath}
\usepackage{graphicx}
\usepackage{float}
\usepackage{tabularx}
\usepackage{color}
\usepackage[table]{xcolor}
\usepackage{parskip}
\usepackage{xfrac}
\usepackage{lineno}
\usepackage{multirow}
\usepackage{setspace}

\begin{document}
\title{Low mechanical loss TiO$_2$:GeO$_2$ coatings for reduced thermal noise in Gravitational Wave Interferometers}

\date{\today}

\author{Gabriele Vajente}\email{vajente@caltech.edu} \affiliation{LIGO Laboratory , California Institute of Technology, Pasadena, CA 91125 USA}
\author{Le Yang} \affiliation{Department of Chemistry, Colorado State University, Fort Collins, CO 80523, USA}
\author{Aaron Davenport} \affiliation{Department of Electrical and Computer Engineering, Colorado State University, Fort Collins, CO 80523, USA}

\author{Mariana Fazio} \affiliation{Department of Electrical and Computer Engineering, Colorado State University, Fort Collins, CO 80523, USA}
\author{Alena Ananyeva} \affiliation{LIGO Laboratory , California Institute of Technology, Pasadena, CA 91125 USA}
\author{Liyuan Zhang} \affiliation{LIGO Laboratory , California Institute of Technology, Pasadena, CA 91125 USA}
\author{Garilynn Billingsley} \affiliation{LIGO Laboratory , California Institute of Technology, Pasadena, CA 91125 USA}

\author{Kiran Prasai}
\affiliation{Edward L. Ginzton Laboratory, Stanford University, Stanford, CA 94305, USA}

\author{Ashot Markosyan}
\affiliation{Edward L. Ginzton Laboratory, Stanford University, Stanford, CA 94305, USA}

\author{Riccardo Bassiri}
\affiliation{Edward L. Ginzton Laboratory, Stanford University, Stanford, CA 94305, USA}

\author{Martin M. Fejer}
\affiliation{Edward L. Ginzton Laboratory, Stanford University, Stanford, CA 94305, USA}

\author{Martin Chicoine} 
\affiliation{D\'epartement de Physique, Universit\'e de Montr\'eal, Montr\'eal, Qu\'ebec, Canada}

\author{Fran\c cois Schiettekatte} 
\affiliation{D\'epartement de Physique, Universit\'e de Montr\'eal, Montr\'eal, Qu\'ebec, Canada}

\author{Carmen S. Menoni} 
\affiliation{Department of Chemistry, Colorado State University, Fort Collins, CO 80523, USA} \affiliation{Department of Electrical and Computer Engineering, Colorado State University, Fort Collins, CO 80523, USA} 

\begin{abstract}
The sensitivity of current and planned gravitational wave interferometric detectors is limited, in the most critical frequency region around 100 Hz, by a combination of quantum noise and thermal noise. The latter is dominated by Brownian noise: thermal motion originating from the elastic energy dissipation in the dielectric coatings used in the interferometer mirrors. The energy dissipation is a material property characterized by the mechanical loss angle. We have identified mixtures of titanium dioxide (TiO$_2$) and germanium dioxide (GeO$_2$) that show internal dissipations at a level of $1 \times 10^{-4}$, low enough to provide almost a factor of two improvement on the level of Brownian noise with respect to the state-of-the-art materials. We show that by using a mixture of 44\% TiO$_2$ and 56\% GeO$_2$ in the high refractive index layers of the interferometer mirrors, it would be possible to achieve a thermal noise level in line with the design requirements. These results are a crucial step forward to produce the mirrors needed to meet the thermal noise requirements for the planned upgrades of the Advanced LIGO and Virgo detectors.
\end{abstract}

\maketitle

Gravitational wave (GW) detectors are highly sensitive instruments that measure the very small distance changes produced by signals of astrophysical origin \cite{PhysRevLett.116.061102, PhysRevX.9.031040}. The current generation of GW detectors are km-scale laser interferometers \cite{TheLIGOScientific:2014jea, TheVirgo:2014hva, Akutsu:2018axf, Vajente:2019a} with several hundreds of kW of circulating power in the Fabry-Perot arm cavities. The test-mass mirrors are made of high-purity fused silica substrates, coated with high-reflectivity multilayer dielectric thin-film stacks \cite{Steinlechner:2018crl}, composed of multiple pairs of high and low refractive index metal oxide layers, making a Bragg reflector structure. 

The sensitivity of the current detectors \cite{PhysRevD.93.112004, PhysRevD.102.062003} is limited by  a combination of laser quantum noise \cite{PhysRevD.64.042006} and displacement noise generated by the Brownian motion of the coatings \cite{BRAGINSKY2003244, PhysRevD.57.659}. Therefore, to increase the astrophysical reach of future detectors, it is crucial to reduce coating Brownian noise. This in turn requires reducing the elastic energy dissipation in the thin film materials composing the coatings \cite{PhysRev.86.702, PhysRevD.57.659}. The power spectral density of Brownian noise at a frequency $f$ is a complex function of the properties of the materials used in the coatings \cite{PhysRevD.87.082001, Fejer2021}. An approximate expression, assuming equal bulk and shear loss angles, is given by (see \cite{Fejer2021} and supplemental material \cite{supplemental}):

\begin{eqnarray}\label{eq:brownian}
S_{\mathrm{B}}(f) = \frac{2k_B T d}{\pi^2 w^2 f} && \left[ \left< \frac{Y}{1-\nu^2} \phi \right>\frac{(1+\nu_S)^2(1-2\nu_S)^2}{Y_S^2}   \right.  \nonumber \\ 
   && +  \left.\left<  \frac{(1+\nu)(1-2\nu)}{(1-\nu) Y}\phi \right> \right]
\end{eqnarray}

where $k_B$ is the Boltzmann's constant, $T$ is the ambient temperature, $w$ is the radius of the laser beam probing the mirror motion, $d$ is the total thickness of the coating, $Y_S$ and $\nu_S$ are the Young's modulus and Poisson ratio of the substrate. The angular bracket expression $\left< x\right>$ indicates the \emph{effective medium} average \cite{Backus_1962, Fejer2021} of the material property $x$ through the stack, weighted by the physical thickness of the layers. The relevant properties of the coating materials are the Young's moduli $Y$, the Poisson ratios $\nu$ and the loss angles $\phi = \mathrm{Im}(Y) / \mathrm{Re}(Y)$. 

The coatings used in the current Advanced LIGO mirrors are composed of alternating layers of amorphous SiO$_2$ of low refractive index $n_{\mathrm{SiO}_2}$ = 1.45 at 1064 nm, and TiO$_2$:Ta$_2$O$_5$ of high refractive index $n_{\mathrm{TiO}_2:\mathrm{Ta}_2\mathrm{O}_5}$ = 2.10 at 1064 nm \cite{Granata_2020, Amato_2019}. The TiO$_2$:Ta$_2$O$_5$ layers have a loss angle much larger than the SiO$_2$ layers ($3-4\times 10^{-4}$ \cite{PhysRevD.98.122001, amato2021} compared to $\sim 2 \times 10^{-5}$ \cite{Granata_2020}) and therefore they dominate in the contribution to the coating Brownian noise.

The goal for the next upgrade to the LIGO detectors, called Advanced LIGO+ \cite{PhysRevD.91.062005, is-whitepaper} is a reduction of the coating noise by about a factor of two, with a target Brownian noise of $S_{\mathrm{B}}^{1/2} = 6.6 \times 10^{-21} \mathrm{m} / \sqrt{\mathrm{Hz}}$ at a frequency of 100 Hz. The SiO$_2$ layers can already be produced with low enough mechanical loss angle \cite{Granata_2020}, so the main focus of the current research is on improving the high refractive index material. Several different approaches have been investigated, including deposition at elevated substrate temperatures \cite{PhysRevLett.113.025503, Vajente_2018} and with assist ion bombardment \cite{yang2019investigation,yang2020modifications}, doping and nanolayering of Ta$_2$O$_5$ \cite{yang2020structural, fazio2020structure, flaminio2010study, zirconia-paper}, and the use of nitrides \cite{Amato_2018, Granata:20}. Here we report results on amorphous oxide coatings based on mixtures of GeO$_2$ and TiO$_2$. 

The initial motivation to investigate coatings based on GeO$_2$ was the discovery of a correlation between the room-temperature mechanical loss angle and the fraction of edge-sharing versus corner-sharing polyhedra in the medium-range order, as reported in \cite{PhysRevLett.123.045501} for ZrO$_2$:Ta$_2$O$_5$. SiO$_2$ also has a prevalence of corner-sharing, and is the amorphous material that exhibits the lowest known room-temperature loss angle in the acoustic frequency range \cite{PENN20063, Ageev_2004, Granata_2020}. Additionally, the mechanical loss angle of GeO$_2$ at low temperatures \cite{PhysRevB.52.7179} ($\lesssim 100$ K) exhibits a peak similar to the one found in SiO$_2$ \cite{Topp1996, Martin_2014}. In recent experiments on GeO$_2$ \cite{yang2021enhanced}, we confirmed that the atomic packing can be altered to improve medium range order by annealing and high temperature deposition. Similar correlations were found for different oxides by other groups \cite{PhysRevB.50.118, doi:10.1063/1.4890958, PhysRevMaterials.2.053607, Amato2020, doi:10.1116/1.5122661}.

From the optical perspective, however, GeO$_2$  has a refractive index $n = 1.60$ at $1064$ nm that makes it unsuitable for use in a high reflector design when combined with SiO$_2$, as 138 layers would be needed to achieve the required high reflectivity for the test-mass mirrors. The increase in the total thickness of a GeO$_2$/SiO$_2$ reflector would balance out the reduced mechanical loss, with no net improvement in the coating Brownian noise. To increase the refractive index at the laser wavelength of 1064 nm, GeO$_2$ was co-deposited with TiO$_2$ with different cation concentrations. 

\begin{figure}[t] 
\begin{center}
\includegraphics[width=\columnwidth]{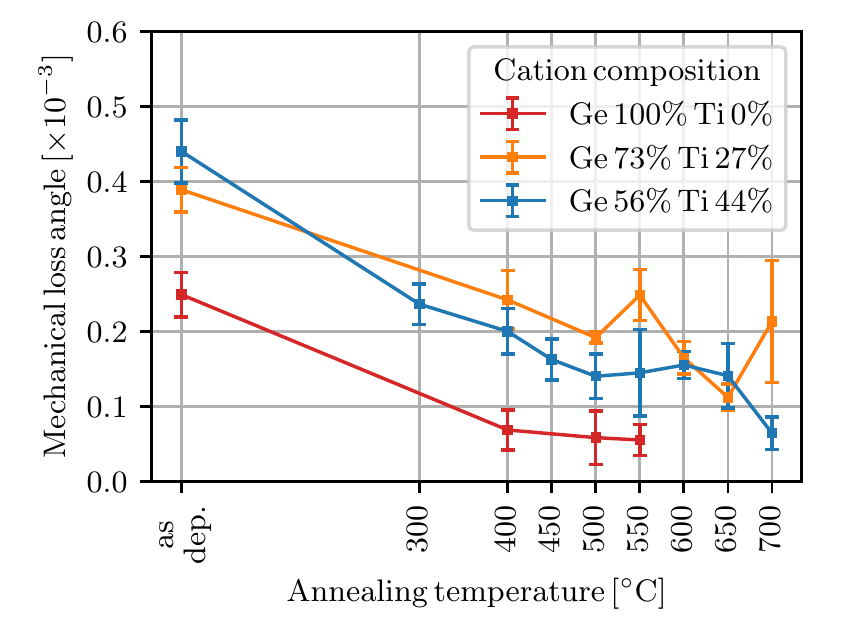} 
\end{center}
\caption{\label{fig:loss-Y} Measured loss angle of TiO$_2$:GeO$_2$, as deposited and after 10-hours-long annealing in air, at increasing temperatures. Different color lines correspond to the cation composition listed in the legend. Only one sample for each concentration is shown here for simplicity. Other samples showed equal values within the error bars.}
\end{figure}

\begin{figure}[t] 
\begin{center}
\includegraphics[width=\columnwidth]{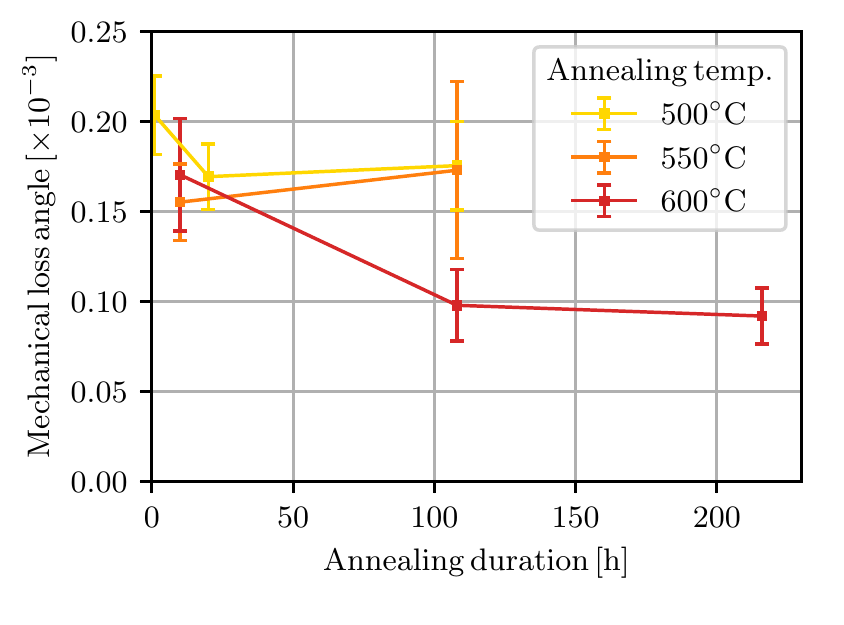} 
\end{center}
\caption{\label{fig:loss-duration} Effect of the annealing duration on the measured loss angle for the 44\% TiO$_2$:GeO$_2$ film.}
\end{figure}

Thin films of TiO$_2$:GeO$_2$ with Ti cation concentration of 0\%, 27\%, and 44\%, were deposited by ion beam sputtering using a biased target deposition system \cite{zhurin2000biased}, that allowed convenient tuning of the mixture composition by adjusting the length of the pulses biasing the metallic Ti and Ge targets.

The cation concentration, oxygen stoichiometry and atomic areal density of the films were determined by Rutherford backscattering spectrometry (RBS) \cite{RBS}. The thickness and refractive index were obtained from spectroscopic ellipsometry. The mass density was computed from the RBS and ellipsometry measurements. The absorption loss at the wavelength of 1064 nm was assessed from photo-thermal common-path interferometry \cite{alexandrovski2009photothermal}. The thin films were annealed in air, as annealing has been shown to reduce absorption loss and room-temperature mechanical loss angle in amorphous oxides \cite{10.1117/12.618288, abernathy2018overview}. Grazing incidence x-ray diffraction shows all mixture films are amorphous upon annealing at 600$^{\circ}$C for 10 and 108 hours, and show signs of crystallization when annealed at higher temperatures. The pure GeO$_2$ film remained amorphous up to 550$^\circ$C.

For the Ti cation concentration of 44\%, the refractive index at 1064 nm is $n_{\mathrm{TiO}_2:\mathrm{Ge}\mathrm{O}_2}$ = 1.88. The absorption loss normalized to a quarter-wavelength (QWL) thick single layer (141 nm) is 2.3 $\pm$ 0.1 ppm after annealing at 600$^\circ$C. The absorption loss of pure GeO$_2$ after annealing at 500$^{\circ}$C is below 1 ppm at $\lambda$ = 1064 nm, showing the potential for improved absorption in the mixture. The deposition parameters are being optimized to achieve even lower optical absorption in TiO$_2$:GeO$_2$, to meet the Advanced LIGO+ requirements of less than 1 ppm \cite{is-whitepaper} for a full mirror coating. More details on the structural and optical characterizations are available in the supplemental material \cite{supplemental}.

The thin films were also deposited with the same procedure on 75-mm-diameter, 1-mm-thick silica disks, to measure the material's elastic properties. The disk acts as a resonator: about 20 modes between 1 kHz and 30 kHz can be measured in a Gentle Nodal Suspension \cite{doi:10.1063/1.3124800, doi:10.1063/1.4990036} to obtain their precise frequency and decay time. After the thin film is deposited on the substrate, the resonant frequencies are shifted by amounts depending on the film properties, allowing an estimation of the Young's modulus and Poisson ratio \cite{Granata_2020, PhysRevD.101.042004}. The decay times of the modes of the coated substrates are significantly shorter than for the bare substrate, due to the elastic energy dissipation in the film. Using the measured elastic properties of the film material, one can compute the fraction of elastic energy in the film for each resonant mode and use it to extract the loss angle $\phi$ of the thin film material \cite{PhysRevD.89.092004}. 

For a homogeneous amorphous material, the relation between stress and strain in the elastic regime can be described in terms of two elastic moduli, for example bulk $K$ and shear $\mu$ moduli \cite{landau1984}. Similarly, the internal energy dissipation in the material should be described in terms of two loss angles $\phi_K = \mathrm{Im}(K) / \mathrm{Re}(K)$ and $\phi_\mu = \mathrm{Im}(\mu) / \mathrm{Re}(\mu)$. There is no physical reason to assume the two loss angles to be equal, and we shall show in the following that they are indeed significantly different for TiO$_2$:GeO$_2$. The layered structure of the stack implies that a description in terms of an equivalent isotropic material is not accurate, since the bulk and shear energy distribution in the layers is different in the case of the ring-down measurements and in the Brownian noise case. While the expression in equation \ref{eq:brownian} assumes equal loss angles, a more precise expression, including the distinction between bulk and shear properties for all layers, is described in the supplemental material \cite{supplemental}, and is needed to correctly account for the multilayer structure and the different materials. 

However, in the initial exploration of the effect of composition and annealing schedule, we relied on the commonly used description with frequency independent equal bulk and shear loss angles \cite{Harry_2002, martin_reid_2012, Granata_2020}. The more detailed analysis of the best candidate material, described later, supports the use of this simplification for survey purposes, since our measurements are more sensitive to the shear than the bulk loss angle, and the former is found to be almost frequency independent. With this approach, figure \ref{fig:loss-Y} shows the measured loss angle for pure GeO$_2$  and the two concentrations of TiO$_2$ and GeO$_2$ studied in detail here. The most promising results are from a mixture of 44\% TiO$_2$ and 56\% GeO$_2$. The mechanical loss of amorphous oxides typically decreases with increasing annealing temperature and time. We observed rapid crystallization at 700$^\circ$C, and therefore explored the effect of annealing duration on the loss angle. We tested heat treatments of 1, 10, 20, 108 and 216 hours in total, for temperatures of 500, 550 and 600$^\circ$C. Figure \ref{fig:loss-duration} shows the effect of annealing time on the loss angle of the 44\% TiO$_2$:GeO$_2$ mixture. It was found that extended annealing at lower temperatures produces little improvement. Instead, after annealing at 600$^\circ$C for 108 hours, the loss angle is reduced to $(0.96 \pm 0.18) \times 10^{-4}$, and the film is still amorphous. Among those tested in our work, this TiO$_2$:GeO$_2$ mixture is the most promising high-index material for low Brownian noise Advanced LIGO+ mirrors, though further characterization of mixtures with other Ti/Ge ratios in this range is planned to find the optimum.

\begin{table}[tb]
\centering
\begin{tabular}{ll}
\textbf{SiO$_2$ property} &\textbf{Value} \\
\hline
Refr. index at 1064 nm & $1.45 \pm 0.01$ \\
Young's modulus  & $73.2 \pm 0.6$ GPa  \\
Poisson ratio & $0.11 \pm 0.07$ \\[2mm]
Loss angle & $\phi_K = \phi_\mu = \left(2.6^{+0.5}_{-0.6} \right) \times 10^{-5}$\\ \\
\textbf{TiO$_2$:GeO$_2$ property} &\textbf{Value} \\
\hline
Cation conc. Ti/(Ti+Ge) & $44.6 \pm 0.3$ \%  \\
Refr. index at 1064 nm & $1.88 \pm 0.01$ \\
Optical abs. for a QWL& 2.3 $\pm$ 0.1 ppm\\
Density & $3690 \pm 100$ kg/m$^3$ \\
Young's modulus  & 91.5 $\pm$ 1.8 GPa \\
Poisson ratio & 0.25 $\pm$ 0.07 \\[2mm]
\multirow{2}{*}{Bulk Loss angle} & $a_K = \left(22.0^{+10.6}_{-12.5}\right) \times 10^{-5}$ \\[0.5mm]
&$m_K = 1.04^{+0.40}_{-0.36}$\\[2mm]
\multirow{2}{*}{Shear Loss angle} & $a_\mu = \left(8.4^{+2.9}_{-4.0}\right) \times 10^{-5}$ \\[0.5mm]
& $m_\mu = -0.06^{+0.15}_{-0.30}$\\
\end{tabular} \caption{Measured parameters for  TiO$_2$:GeO$_2$ and SiO$_2$,  after annealing at 600$^\circ$C for 108 hours. The loss angle model for  TiO$_2$:GeO$_2$ is $\phi(f) = a \cdot (f/10\,\mathrm{kHz})^{m}$. Uncertainties describe the 90\% confidence intervals.
\label{tab:parameters}}
\end{table}

As a first step toward the production of a full high-reflectivity coating, and to better characterize this new material, we deposited single layers of SiO$_2$ and TiO$_2$:GeO$_2$, as well as a stack of 5 QWL layers of TiO$_2$:GeO$_2$ alternated with 5 layers of SiO$_2$, and 20 layers of TiO$_2$:GeO$_2$ alternated with 20 layers of SiO$_2$. The depositions were performed using a commercial Spector Ion Beam Sputtering system that can produce films with better optical quality \cite{yang2019investigation} than the biased target system used for the initial parameter exploration.  At the laser wavelength of 1064 nm, the transmission of the 40-layer structure was 190 ppm and the optical absorption was measured to be 3.1 ppm after annealing. We also measured the Young's modulus and loss angle of the stacks. However, since the multilayers structure is not isotropic, a description in terms of Y and $\nu$ is only approximate. Nevertheless, the two stacks were found to have the same Young's modulus, $78.0 \pm 1.3$ GPa, and the same loss angle, $(5.5 \pm 0.7) \times 10^{-5}$ after annealing at 600$^\circ$C for 108 h. This is an indication that there is no evidence of any systematic error in the measurements due to the thickness of the coatings. At an approximation level consistent with assuming equal bulk and shear loss angles, one can compute the expected value for the stack by averaging the single material values as $\bar \phi = \left< Y \phi\right> / \left<Y\right> = (6.4 \pm 1.7) \times 10^{-5}$. Therefore there is no indication of excess loss due to interfaces \cite{PhysRevD.93.012007}.

\begin{figure}[t!] 
\begin{center}
\includegraphics[width=\columnwidth]{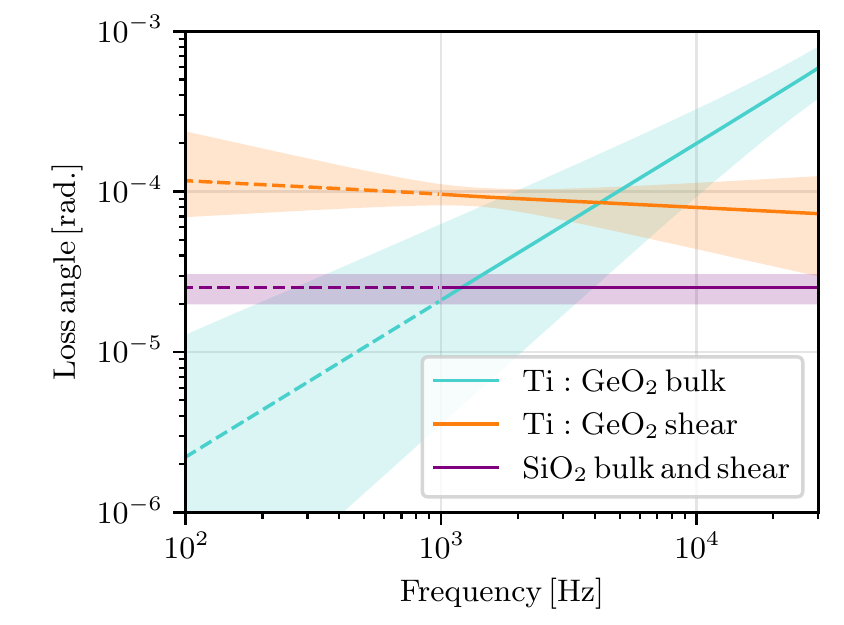} 
\end{center}
\caption{\label{fig:loss-angle} Estimated bulk and shear loss angles as a function of frequency. The solid lines indicate the range of frequencies where the loss angles were measured, while the dashed lines are extrapolations to the lower frequency range. The shaded regions show the 90\% confidence intervals of the estimates.}
\end{figure}

A correct description of the Brownian noise in a multilayer stack must take into account the bulk and shear moduli and loss angles of the individual materials. The resonant modes of the coated disk store different fractions of bulk and shear energy in the film, and therefore it is possible to extract the bulk and shear loss angles from the measurements \cite{ABERNATHY20182282, PhysRevD.101.042004}. We model a single isotropic layer with known thickness and density as measured by ellipsometry and RBS, and with Young's modulus, Poisson ratio and bulk and shear loss angles as free parameters. For each sample, the measurement data set consists of the frequency shifts due to the coating, and the reduction in the decay time, due to the energy dissipation in the coating, for each of the measurable modes. We used a Markov chain Monte Carlo Bayesian Analysis \cite{PhysRevD.101.042004, BDA, emcee3} to find the probability distribution of the model parameters given the data. We considered either different bulk and shear loss angles or equal loss angles, and three possible frequency dependencies: constant, linear or power law, for a total of six different loss models. The Bayesian analysis allows us to compute the relative likelihood of each model given the data. The best model for the TiO$_2$:GeO$_2$ film is a power law with different bulk and shear loss angles, while for the SiO$_2$ film it is a constant single loss angle, as shown in figure \ref{fig:loss-angle}. It is worth noting that the bulk loss angle for the TiO$_2$:GeO$_2$ film shows a rather steep frequency dependence. The second most likely model for this material is the one with a linear frequency dependence. The bulk loss angle does not show a frequency dependency as steep as in the power law case, but it is still predicted to be significantly smaller than the shear loss angle at low frequency. The measured value of the loss angle for SiO$_2$ is compatible with values reported in the literature \cite{Granata_2020}. Table \ref{tab:parameters} summarizes all the measured material properties. More details on the analysis and the results are in the supplemental material \cite{supplemental}.

\begin{figure}[t!] 
\begin{center}
\includegraphics[width=\columnwidth]{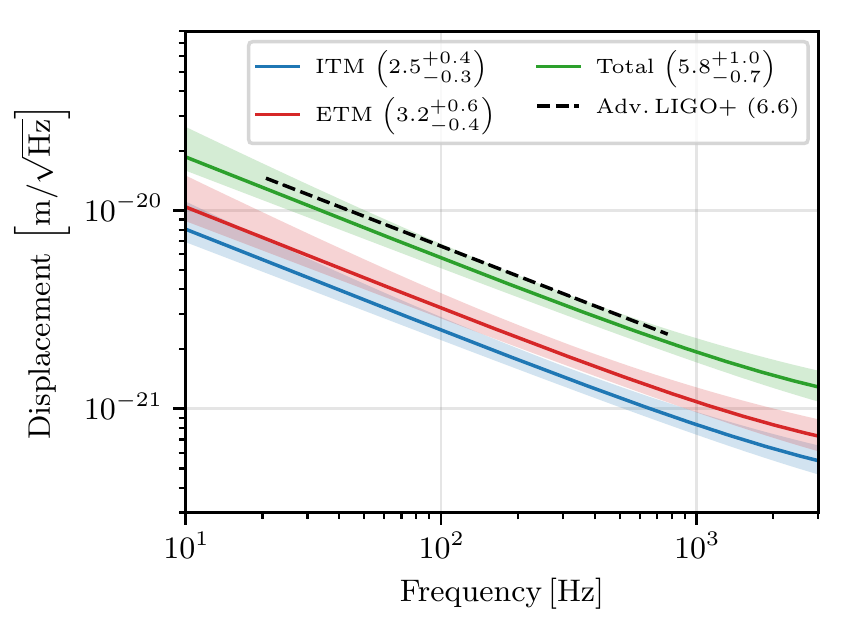} 
\end{center}
\caption{\label{fig:brownian-noise} Estimated Brownian noise for the Advanced LIGO+ interferometer. The red and blue traces show the contribution of a single ITM and ETM, while the green trace shows the total for all four test masses. The numbers in the legend give the Brownian noise level at 100 Hz, in units of $10^{-21} \mathrm{m}/\sqrt{\mathrm{Hz}}$. The shaded regions correspond to the 90\% confidence intervals of the estimates. The dashed black line shows the design target for Advanced LIGO+.}
\end{figure}

The transmission requirements for the Advanced LIGO+ test masses are similar to those for Advanced LIGO: the input mirror test masses (ITM) should have a transmission of 1.4\% and the end test masses (ETM) of 5 ppm \cite{TheLIGOScientific:2014jea}. Given the measured refractive indexes, the ITM stack is composed of 11 layers of 106 nm of TiO$_2$:GeO$_2$ alternated with 11 layers of 228 nm of SiO$_2$, while the ETM stack is composed of 26 layers of 123 nm of TiO$_2$:GeO$_2$ and 26 layers of 207 nm of SiO$_2$. Both structures are capped with a half-wavelength-thick SiO$_2$ layer.

The Brownian noise for such mirrors can be computed using the effective medium approach described in the supplemental material \cite{supplemental}, which has been checked to provide results within a few percent of other published formulas \cite{PhysRevD.87.082001, PhysRevD.91.042002}. The results are shown in figure \ref{fig:brownian-noise}. The noise is compliant with the design requirement for Advanced LIGO+, reaching $\left(5.8^{+1.0}_{-0.7}\right) \times 10^{-21}$ m$/\sqrt{\mathrm{Hz}}$ at 100 Hz. It is worth noting that this result does not depend strongly on the steep frequency dependency predicted for the bulk loss angle of TiO$_2$:GeO$_2$. If we use the second most probable model, with a less steep frequency dependency, we obtain $\left(6.2^{+1.5}_{-0.8}\right) \times 10^{-21}$ m$/\sqrt{\mathrm{Hz}}$ at 100 Hz, still compatible with the Advanced LIGO+ design requirement. Preparation of samples on disks with lower resonant frequencies to better constrain these estimates is underway.

In summary, we demonstrated that a mixture of 44\% TiO$_2$ and 56\% GeO$_2$ offers excellent optical quality and low mechanical loss angle, making it a promising material to be used as high-index layer in the test mass coatings of the Advanced LIGO+ interferometric GW detectors. We analyzed the internal energy dissipation of this novel material in terms of bulk and shear loss angles, and used the results to design multilayer high reflectivity stacks for the Advanced LIGO+ mirrors. The Brownian noise achievable with TiO$_2$:GeO$_2$ / SiO$_2$ based mirrors reaches a level compliant with the Advanced LIGO+ design requirements.

Studies are on-going to further improve mechanical and optical absorption losses by changing  deposition parameters, mixture and annealing schedule, and to characterize the scattering properties of the multilayer stacks. We are also planning to directly measure the Brownian noise of optimized high reflection mirrors designed for Advanced LIGO+ \cite{PhysRevD.98.122001}, to confirm the noise prediction. 

\subsubsection{Acknowledgement}

This work is supported by the National Science Foundation (NSF) LIGO program through grants No. 1710957 and 1708010. We also acknowledge the support of the LSC Center for Coatings Research, jointly funded by the National Science Foundation and the Gordon and Betty Moore Foundation (GBMF). K. P., A. M., R. B. and M. M. F. are grateful for support through NSF awards PHY-1707866, PHY-1708175, PHY-2011571, PHY-2011706, and GBMF Grant No. 6793. The work carried out at U. Montreal benefited from the support of the NSERC, the CFI and the FRQNT through the RQMP.


\begin{thebibliography}{10}

\bibitem{PhysRevLett.116.061102}
B.~P. Abbott et~al.
\newblock Observation of gravitational waves from a binary black hole merger.
\newblock {\em Phys. Rev. Lett.}, 116:061102, Feb 2016.

\bibitem{PhysRevX.9.031040}
B.~P. Abbott et~al.
\newblock {GWTC-1: A Gravitational-Wave Transient Catalog of Compact Binary
  Mergers Observed by LIGO and Virgo during the First and Second Observing
  Runs}.
\newblock {\em Phys. Rev. X}, 9:031040, Sep 2019.

\bibitem{TheLIGOScientific:2014jea}
J.~Aasi et~al.
\newblock {Advanced LIGO}.
\newblock {\em Class. Quantum Grav.}, 32:074001, 2015.

\bibitem{TheVirgo:2014hva}
F.~Acernese et~al.
\newblock {Advanced Virgo: a second-generation interferometric gravitational
  wave detector}.
\newblock {\em Class. Quantum Grav.}, 32(2):024001, 2015.

\bibitem{Akutsu:2018axf}
T.~Akutsu et~al.
\newblock {KAGRA: 2.5 Generation Interferometric Gravitational Wave Detector}.
\newblock {\em Nat. Astron.}, 3(1):35--40, 2019.

\bibitem{Vajente:2019a}
Gabriele Vajente, Eric~K. Gustafson, and David~H. Reitze.
\newblock {Chapter Three - Precision interferometry for gravitational wave
  detection: Current status and future trends}.
\newblock volume~68 of {\em Advances In Atomic, Molecular, and Optical
  Physics}, pages 75 -- 148. Academic Press, 2019.

\bibitem{Steinlechner:2018crl}
J.~Steinlechner.
\newblock {Development of mirror coatings for gravitational-wave detectors}.
\newblock {\em Phil. Trans. Roy. Soc. Lond. A}, 376(2120):20170282, 2018.

\bibitem{PhysRevD.93.112004}
D.~V. Martynov et~al.
\newblock {Sensitivity of the Advanced LIGO detectors at the beginning of
  gravitational wave astronomy}.
\newblock {\em Phys. Rev. D}, 93:112004, Jun 2016.

\bibitem{PhysRevD.102.062003}
A.~Buikema et~al.
\newblock {Sensitivity and performance of the Advanced LIGO detectors in the
  third observing run}.
\newblock {\em Phys. Rev. D}, 102:062003, Sep 2020.

\bibitem{PhysRevD.64.042006}
Alessandra Buonanno and Yanbei Chen.
\newblock {Quantum noise in second generation, signal-recycled laser
  interferometric gravitational-wave detectors}.
\newblock {\em Phys. Rev. D}, 64:042006, Jul 2001.

\bibitem{BRAGINSKY2003244}
V.B. Braginsky and S.P. Vyatchanin.
\newblock {Thermodynamical fluctuations in optical mirror coatings}.
\newblock {\em Physics Letters A}, 312(3):244 -- 255, 2003.

\bibitem{PhysRevD.57.659}
Yu. Levin.
\newblock {Internal thermal noise in the LIGO test masses: A direct approach}.
\newblock {\em Phys. Rev. D}, 57:659--663, Jan 1998.

\bibitem{PhysRev.86.702}
Herbert~B. Callen and Richard~F. Greene.
\newblock {On a Theorem of Irreversible Thermodynamics}.
\newblock {\em Phys. Rev.}, 86:702--710, Jun 1952.

\bibitem{PhysRevD.87.082001}
Ting Hong, Huan Yang, Eric~K. Gustafson, Rana~X. Adhikari, and Yanbei Chen.
\newblock {Brownian thermal noise in multilayer coated mirrors}.
\newblock {\em Phys. Rev. D}, 87:082001, Apr 2013.

\bibitem{Fejer2021}
M.~M. Fejer.
\newblock {Effective Medium Description of Multilayer Coatings}.
\newblock {\em LIGO Document}, T2100186, 2021.

\bibitem{supplemental}
See supplemental material at [url will be inserted by publisher].

\bibitem{Backus_1962}
George~E. Backus.
\newblock Long-wave elastic anisotropy produced by horizontal layering.
\newblock {\em Journal of Geophysical Research (1896-1977)}, 67(11):4427--4440,
  1962.

\bibitem{Granata_2020}
M~Granata, A~Amato, L~Balzarini, M~Canepa, J~Degallaix, D~Forest, V~Dolique,
  L~Mereni, C~Michel, L~Pinard, B~Sassolas, J~Teillon, and G~Cagnoli.
\newblock {Amorphous optical coatings of present gravitational-wave
  interferometers}.
\newblock {\em Classical and Quantum Gravity}, 37(9):095004, apr 2020.

\bibitem{Amato_2019}
Alex Amato, Silvana Terreni, Vincent Dolique, Dani{\`{e}}le Forest, Gianluca
  Gemme, Massimo Granata, Lorenzo Mereni, Christophe Michel, Laurent Pinard,
  Benoit Sassolas, Julien Teillon, Gianpietro Cagnoli, and Maurizio Canepa.
\newblock {Optical properties of high-quality oxide coating materials used in
  gravitational-wave advanced detectors}.
\newblock {\em Journal of Physics: Materials}, 2(3):035004, jun 2019.

\bibitem{PhysRevD.98.122001}
S.~Gras and M.~Evans.
\newblock {Direct measurement of coating thermal noise in optical resonators}.
\newblock {\em Phys. Rev. D}, 98:122001, Dec 2018.

\bibitem{amato2021}
A~Amato et~al.
\newblock {Optical and mechanical properties of ion-beam-sputtered Nb$_2$O$_5$
  and TiO$_2$-Nb$_2$O$_5$ thin films for gravitational-wave interferometers and
  an improved measurement of coating thermal noise in Advanced LIGO}.
\newblock {\em Accepted in Phys. Rev. D}, 2021.

\bibitem{PhysRevD.91.062005}
John Miller, Lisa Barsotti, Salvatore Vitale, Peter Fritschel, Matthew Evans,
  and Daniel Sigg.
\newblock {Prospects for doubling the range of Advanced LIGO}.
\newblock {\em Phys. Rev. D}, 91:062005, Mar 2015.

\bibitem{is-whitepaper}
The LIGO~Scientific Collaboration.
\newblock {Instrument Science White Paper 2020}.
\newblock {\em LIGO document}, T2000407, 2020.

\bibitem{PhysRevLett.113.025503}
Xiao Liu, Daniel~R. Queen, Thomas~H. Metcalf, Julie~E. Karel, and Frances
  Hellman.
\newblock {Hydrogen-Free Amorphous Silicon with No Tunneling States}.
\newblock {\em Phys. Rev. Lett.}, 113:025503, Jul 2014.

\bibitem{Vajente_2018}
G~Vajente, R~Birney, A~Ananyeva, S~Angelova, R~Asselin, B~Baloukas, R~Bassiri,
  G~Billingsley, M~M Fejer, D~Gibson, L~J Godbout, E~Gustafson, A~Heptonstall,
  J~Hough, S~MacFoy, A~Markosyan, I~W Martin, L~Martinu, P~G Murray, S~Penn,
  S~Roorda, S~Rowan, F~Schiettekatte, R~Shink, C~Torrie, D~Vine, S~Reid, and
  R~X Adhikari.
\newblock {Effect of elevated substrate temperature deposition on the
  mechanical losses in tantala thin film coatings}.
\newblock {\em Classical and Quantum Gravity}, 35(7):075001, feb 2018.

\bibitem{yang2019investigation}
Le~Yang, Emmett Randel, Gabriele Vajente, Alena Ananyeva, Eric Gustafson, Ashot
  Markosyan, Riccardo Bassiri, Martin~M Fejer, and Carmen~S Menoni.
\newblock {Investigation of effects of assisted ion bombardment on mechanical
  loss of sputtered tantala thin films for gravitational wave interferometers}.
\newblock {\em Physical Review D}, 100(12):122004, 2019.

\bibitem{yang2020modifications}
Le~Yang, Emmett Randel, Gabriele Vajente, Alena Ananyeva, Eric Gustafson, Ashot
  Markosyan, Riccardo Bassiri, Martin Fejer, and Carmen Menoni.
\newblock {Modifications of ion beam sputtered tantala thin films by secondary
  argon and oxygen bombardment}.
\newblock {\em Applied Optics}, 59(5):A150--A154, 2020.

\bibitem{yang2020structural}
Le~Yang, Mariana Fazio, Gabriele Vajente, Alena Ananyeva, GariLynn Billingsley,
  Ashot Markosyan, Riccardo Bassiri, Martin~M Fejer, and Carmen~S Menoni.
\newblock {Structural Evolution that Affects the Room-Temperature Internal
  Friction of Binary Oxide Nanolaminates: Implications for Ultrastable Optical
  Cavities}.
\newblock {\em ACS Applied Nano Materials}, 2020.

\bibitem{fazio2020structure}
Mariana~A Fazio, Gabriele Vajente, Alena Ananyeva, Ashot Markosyan, Riccardo
  Bassiri, Martin~M Fejer, and Carmen~S Menoni.
\newblock {Structure and morphology of low mechanical loss TiO 2-doped Ta 2 O
  5}.
\newblock {\em Optical Materials Express}, 10(7):1687--1703, 2020.

\bibitem{flaminio2010study}
Raffaele Flaminio, Janyce Franc, Christine Michel, Nazario Morgado, Laurent
  Pinard, and Benoit Sassolas.
\newblock {A study of coating mechanical and optical losses in view of reducing
  mirror thermal noise in gravitational wave detectors}.
\newblock {\em Classical and Quantum Gravity}, 27(8):084030, 2010.

\bibitem{zirconia-paper}
M.~Abernathy et~al.
\newblock {Exploration of co-sputtered Ta$_2$O$_5$-ZrO$_2$ thin films for
  gravitational-wave detectors}.
\newblock {\em Submitted to Classical and Quantum Gravity}, 2021.

\bibitem{Amato_2018}
Alex Amato, Gianpietro Cagnoli, Maurizio Canepa, Elodie Coillet, Jerome
  Degallaix, Vincent Dolique, Daniele Forest, Massimo Granata, Val{\'{e}}rie
  Martinez, Christophe Michel, Laurent Pinard, Benoit Sassolas, and Julien
  Teillon.
\newblock {High-Reflection Coatings for Gravitational-Wave Detectors:State of
  The Art and Future Developments}.
\newblock {\em Journal of Physics: Conference Series}, 957:012006, feb 2018.

\bibitem{Granata:20}
M.~Granata, A.~Amato, G.~Cagnoli, M.~Coulon, J.~Degallaix, D.~Forest,
  L.~Mereni, C.~Michel, L.~Pinard, B.~Sassolas, and J.~Teillon.
\newblock {Progress in the measurement and reduction of thermal noise in
  optical coatings for gravitational-wave detectors}.
\newblock {\em Appl. Opt.}, 59(5):A229--A235, Feb 2020.

\bibitem{PhysRevLett.123.045501}
K.~Prasai, J.~Jiang, A.~Mishkin, B.~Shyam, S.~Angelova, R.~Birney, D.~A.
  Drabold, M.~Fazio, E.~K. Gustafson, G.~Harry, S.~Hoback, J.~Hough,
  C.~L\'evesque, I.~MacLaren, A.~Markosyan, I.~W. Martin, C.~S. Menoni, P.~G.
  Murray, S.~Penn, S.~Reid, R.~Robie, S.~Rowan, F.~Schiettekatte, R.~Shink,
  A.~Turner, G.~Vajente, H-P. Cheng, M.~M. Fejer, A.~Mehta, and R.~Bassiri.
\newblock {High Precision Detection of Change in Intermediate Range Order of
  Amorphous Zirconia-Doped Tantala Thin Films Due to Annealing}.
\newblock {\em Phys. Rev. Lett.}, 123:045501, Jul 2019.

\bibitem{PENN20063}
Steven~D. Penn, Alexander Ageev, Dan Busby, Gregory~M. Harry, Andri~M.
  Gretarsson, Kenji Numata, and Phil Willems.
\newblock {Frequency and surface dependence of the mechanical loss in fused
  silica}.
\newblock {\em Physics Letters A}, 352(1):3 -- 6, 2006.

\bibitem{Ageev_2004}
Alexandr Ageev, Belkis~Cabrera Palmer, Antonio~De Felice, Steven~D Penn, and
  Peter~R Saulson.
\newblock {Very high quality factor measured in annealed fused silica}.
\newblock {\em Classical and Quantum Gravity}, 21(16):3887--3892, jul 2004.

\bibitem{PhysRevB.52.7179}
Sonja Rau, Christian Enss, Siegfried Hunklinger, Peter Neu, and Alois W\"urger.
\newblock {Acoustic properties of oxide glasses at low temperatures}.
\newblock {\em Phys. Rev. B}, 52:7179--7194, Sep 1995.

\bibitem{Topp1996}
K.~A. Topp and David~G. Cahill.
\newblock {Elastic properties of several amorphous solids and disordered
  crystals below 100 K}.
\newblock {\em Zeitschrift f{\"u}r Physik B Condensed Matter}, 101(2):235--245,
  Mar 1996.

\bibitem{Martin_2014}
I~W Martin, R~Nawrodt, K~Craig, C~Schwarz, R~Bassiri, G~Harry, J~Hough, S~Penn,
  S~Reid, R~Robie, and S~Rowan.
\newblock Low temperature mechanical dissipation of an ion-beam sputtered
  silica film.
\newblock {\em Classical and Quantum Gravity}, 31(3):035019, jan 2014.

\bibitem{yang2021enhanced}
Le~Yang, Gabriele Vajente, Mariana Fazio, Alena Ananyeva, GariLynn Billingsley,
  Ashot Markosyan, Riccardo Bassiri, Kiran Prasai, Martin~M. Fejer, and
  Carmen~S. Menoni.
\newblock {Enhanced Medium Range Order in Vapor Deposited Germania Glasses at
  Elevated Temperatures}.
\newblock {\em arXiv preprint arXiv:2102.08526}, 2021.

\bibitem{PhysRevB.50.118}
Wei Jin, Rajiv~K. Kalia, Priya Vashishta, and Jos\'e~P. Rino.
\newblock Structural transformation in densified silica glass: A
  molecular-dynamics study.
\newblock {\em Phys. Rev. B}, 50:118--131, Jul 1994.

\bibitem{doi:10.1063/1.4890958}
Rashid Hamdan, Jonathan~P. Trinastic, and H.~P. Cheng.
\newblock Molecular dynamics study of the mechanical loss in amorphous pure and
  doped silica.
\newblock {\em The Journal of Chemical Physics}, 141(5):054501, 2014.

\bibitem{PhysRevMaterials.2.053607}
Massimo Granata, Elodie Coillet, Val\'erie Martinez, Vincent Dolique, Alex
  Amato, Maurizio Canepa, J\'er\'emie Margueritat, Christine Martinet, Alain
  Mermet, Christophe Michel, Laurent Pinard, Beno\^{\i}t Sassolas, and
  Gianpietro Cagnoli.
\newblock Correlated evolution of structure and mechanical loss of a sputtered
  silica film.
\newblock {\em Phys. Rev. Materials}, 2:053607, May 2018.

\bibitem{Amato2020}
Alex Amato, Silvana Terreni, Massimo Granata, Christophe Michel, Benoit
  Sassolas, Laurent Pinard, Maurizio Canepa, and Gianpietro Cagnoli.
\newblock {Observation of a Correlation Between Internal friction and Urbach
  Energy in Amorphous Oxides Thin Films}.
\newblock {\em Scientific Reports}, 10(1):1670, Feb 2020.

\bibitem{doi:10.1116/1.5122661}
Alex Amato, Silvana Terreni, Massimo Granata, Christophe Michel, Laurent
  Pinard, Gianluca Gemme, Maurizio Canepa, and Gianpietro Cagnoli.
\newblock {Effect of heating treatment and mixture on optical properties of
  coating materials used in gravitational-wave detectors}.
\newblock {\em Journal of Vacuum Science \& Technology B}, 37(6):062913, 2019.

\bibitem{zhurin2000biased}
Zhurin V~V et~al.
\newblock {Biased target deposition}.
\newblock {\em Journal of Vacuum Science \& Technology A: Vacuum, Surfaces, and
  Films}, 18(1):37, 2000.

\bibitem{RBS}
Wei-Kan Chu.
\newblock {\em {Backscattering Spectrometry, 1st Edition}}.
\newblock {Academic Press}, 1978.

\bibitem{alexandrovski2009photothermal}
Alexei Alexandrovski, Martin Fejer, A~Markosian, and Roger Route.
\newblock {Photothermal common-path interferometry (PCI): new developments}.
\newblock In {\em Solid State Lasers XVIII: Technology and Devices}, volume
  7193, page 71930D. International Society for Optics and Photonics, 2009.

\bibitem{10.1117/12.618288}
Roger~P. Netterfield, Mark Gross, Fred~N. Baynes, Katie~L. Green, Gregory~M.
  Harry, Helena Armandula, Sheila Rowan, Jim Hough, David R.~M. Crooks,
  Martin~M. Fejer, Roger Route, and Steven~D. Penn.
\newblock {Low mechanical loss coatings for LIGO optics: progress report}.
\newblock In Michael~L. Fulton and Jennifer D.~T. Kruschwitz, editors, {\em
  Advances in Thin-Film Coatings for Optical Applications II}, volume 5870,
  pages 144 -- 152. International Society for Optics and Photonics, SPIE, 2005.

\bibitem{abernathy2018overview}
Matthew~Robert Abernathy, Xiao Liu, and Thomas~H Metcalf.
\newblock {An overview of research into low internal friction optical coatings
  by the gravitational wave detection community}.
\newblock {\em Materials Research}, 21, 2018.

\bibitem{doi:10.1063/1.3124800}
E.~Cesarini, M.~Lorenzini, E.~Campagna, F.~Martelli, F.~Piergiovanni,
  F.~Vetrano, G.~Losurdo, and G.~Cagnoli.
\newblock {A ?gentle? nodal suspension for measurements of the acoustic
  attenuation in materials}.
\newblock {\em Review of Scientific Instruments}, 80(5):053904, 2009.

\bibitem{doi:10.1063/1.4990036}
G.~Vajente, A.~Ananyeva, G.~Billingsley, E.~Gustafson, A.~Heptonstall,
  E.~Sanchez, and C.~Torrie.
\newblock {A high throughput instrument to measure mechanical losses in thin
  film coatings}.
\newblock {\em Review of Scientific Instruments}, 88(7):073901, 2017.

\bibitem{PhysRevD.101.042004}
Gabriele Vajente, Mariana Fazio, Le~Yang, Anchal Gupta, Alena Ananyeva,
  Garilynn Billinsley, and Carmen~S. Menoni.
\newblock {Method for the experimental measurement of bulk and shear loss
  angles in amorphous thin films}.
\newblock {\em Phys. Rev. D}, 101:042004, Feb 2020.

\bibitem{PhysRevD.89.092004}
Tianjun Li, Felipe~A. Aguilar~Sandoval, Mickael Geitner, Ludovic Bellon,
  Gianpietro Cagnoli, J\'er\^ome Degallaix, Vincent Dolique, Raffaele Flaminio,
  Dani\`ele Forest, Massimo Granata, Christophe Michel, Nazario Morgado, and
  Laurent Pinard.
\newblock {Measurements of mechanical thermal noise and energy dissipation in
  optical dielectric coatings}.
\newblock {\em Phys. Rev. D}, 89:092004, May 2014.

\bibitem{landau1984}
L~D Landau, L~P Pitaevskii, A~M Kosevich, and E~M Lifshitz.
\newblock {\em Theory of Elasticity, 3rd edition}.
\newblock Elsevier, 1986.

\bibitem{Harry_2002}
Gregory~M Harry, Andri~M Gretarsson, Peter~R Saulson, Scott~E Kittelberger,
  Steven~D Penn, William~J Startin, Sheila Rowan, Martin~M Fejer, D~R~M Crooks,
  Gianpietro Cagnoli, Jim Hough, and Norio Nakagawa.
\newblock Thermal noise in interferometric gravitational wave detectors due to
  dielectric optical coatings.
\newblock {\em Classical and Quantum Gravity}, 19(5):897--917, feb 2002.

\bibitem{martin_reid_2012}
Iain Martin and Stuart Reid.
\newblock {\em Coating thermal noise}, page 31?54.
\newblock Cambridge University Press, 2012.

\bibitem{PhysRevD.93.012007}
Massimo Granata, Emeline Saracco, Nazario Morgado, Alix Cajgfinger, Gianpietro
  Cagnoli, J\'er\^ome Degallaix, Vincent Dolique, Dani\`ele Forest, Janyce
  Franc, Christophe Michel, Laurent Pinard, and Raffaele Flaminio.
\newblock {Mechanical loss in state-of-the-art amorphous optical coatings}.
\newblock {\em Phys. Rev. D}, 93:012007, Jan 2016.

\bibitem{ABERNATHY20182282}
Matthew Abernathy, Gregory Harry, Jonathan Newport, Hannah Fair, Maya
  Kinley-Hanlon, Samuel Hickey, Isaac Jiffar, Andri Gretarsson, Steve Penn,
  Riccardo Bassiri, Eric Gustafson, Iain Martin, Sheila Rowan, and Jim Hough.
\newblock Bulk and shear mechanical loss of titania-doped tantala.
\newblock {\em Physics Letters A}, 382(33):2282--2288, 2018.
\newblock Special Issue in memory of Professor V.B. Braginsky.

\bibitem{BDA}
A~Gelman, J~B Carlin, H~S Stern, D~B Dunson, A~Vehtari, and D~B Rubin.
\newblock {\em Bayesian Data Analysis, 3rd edition}.
\newblock Chapman and Hall, 2013.

\bibitem{emcee3}
Daniel {Foreman-Mackey}, David~W. {Hogg}, Dustin {Lang}, and Jonathan
  {Goodman}.
\newblock {emcee: The MCMC Hammer}.
\newblock {\em Publications of the Astronomical Society of the Pacific},
  125(925):306, March 2013.

\bibitem{PhysRevD.91.042002}
William Yam, Slawek Gras, and Matthew Evans.
\newblock Multimaterial coatings with reduced thermal noise.
\newblock {\em Phys. Rev. D}, 91:042002, Feb 2015.

\bibitem{thamaphat2008phase}
Kheamrutai Thamaphat, Pichet Limsuwan, and Boonlaer Ngotawornchai.
\newblock {Phase characterization of TiO2 powder by XRD and TEM}.
\newblock {\em Agriculture and Natural Resources}, 42(5):357--361, 2008.

\bibitem{MayerRBS}
Matej Mayer.
\newblock {SIMNRA, a simulation program for the analysis of NRA, RBS and ERDA}.
\newblock {\em AIP Conference Proceedings}, 475, 1999.

\bibitem{Crooks_2002}
D~R~M Crooks, P~Sneddon, G~Cagnoli, J~Hough, S~Rowan, M~M Fejer, E~Gustafson,
  R~Route, N~Nakagawa, D~Coyne, G~M Harry, and A~M Gretarsson.
\newblock Excess mechanical loss associated with dielectric mirror coatings on
  test masses in interferometric gravitational wave detectors.
\newblock {\em Classical and Quantum Gravity}, 19(15):4229--4229, jul 2002.

\bibitem{COMSOL}
Comsol multiphysics.

\end{thebibliography}
\end{document}